\documentclass[letterpaper]{jpconf}

\usepackage{graphicx}
\usepackage{bm}

\def\be{\begin{equation}}
\def\ee{\end{equation}}
\def\bea{\begin{eqnarray}}
\def\eea{\end{eqnarray}}
\def\bi{\begin{itemize}}
\def\ei{\end{itemize}}

\begin{document}

\title{Local virial relation and velocity anisotropy in self-gravitating system}
\author{Y. Sota$^{1,2}$, O. Iguchi$^{1}$, 
M. Morikawa$^{1}$ and A. Nakamichi$^{3}$}

\address{$^1$ Department of Physics, Ochanomizu University,
	       2-1-1 Ohtuka, Bunkyo, Tokyo 112-8610, Japan}
\address{$^2$ Advanced Research Institute for Science and Engineering, 
               Waseda University, Ohkubo, Shinjuku--ku, Tokyo 169-8555, Japan}
\address{$^3$ Gunma Astronomical Observatory,
               6860-86, Nakayama, Takayama, Agatsuma, Gunma 377-0702,
  Japan}

\begin{abstract}

We investigate the merging process in N-body self-gravitating system from 
the viewpoints of the local virial relation which is the relation 
between the  local kinetic energy and the  local potential. 
We compare both the density 
profile and the phase space density profile in cosmological simulations with 
the critical solutions of collisionless static
state  satisfying the local virial (LV) relation. We 
got the results that the critical solution can explain the characteristic 
density profile with the appropriate value of  anisotropy parameter $\beta \sim 0.5$. 
It can also explain the power law of the phase space
density profile in the outer part of 
a bound state.  
However, it fails in explaining the central low temperature 
part which is connected to the  scale invariant  phase space 
density. It can be well fitted to the critical solution with the higher value of
$\beta \sim 0.75$. These results indicate that the LV  relation
is not compatible with the  scale invariant 
phase space density in cosmological simulation.
\end{abstract}


\section{\label{sec:intro}Introduction}

It is well known that astronomical objects are formed through the merging 
process of small clusters under the homogeneous expanding background. 
It is  also well known that the  bound state after the
merging process  in  cosmological simulations takes rather universal form of a 
cuspy density profile, $\rho \sim r^{-\gamma}$ in the inner part, where $\gamma \approx 1$ \cite{Navarro96}. 
There is another universal character for the virialized state following 
merging process, that is, the scaling law of a phase space density, $\rho/\sigma^3 \sim r^{-\alpha}$ where $\alpha
\approx 1.9$
\cite{Taylor01}. Several models are proposed to explain 
both of these characters.
For example, W.Dehnen and D.E.McLaughlin show that the
critical solution for the static Vlasov equations under the 
scale invariant  phase space density 
 is compatible with NFW density profile \cite{Dehnen05}.

In our previous paper, we examined  cold collapse simulations
and got the result 
that the bound state after a collapse has several remarkable characters \cite{Iguchi04,Sota04,Osamu05}. 
The LV relation is one of such characters. The 
Plummer model which is the analytical solution for polytrope with 
index n=5 satisfies this condition. The Plummer model has the property 
that the velocity dispersion is isotropic.
 However,  the velocity dispersion is not always isotropic for the
quasi-equilibrium state and  is 
characterized with non-zero anisotropy parameter $\beta \left( r \right)$. 
The analytical solution exists even for the non-zero 
constant value of $\beta $ under the condition of the LV relation
\cite{Evans05}. We showed by combining these analytical 
solutions that the bound state after the cold collapse can be described 
quite well by the solution with the LV condition \cite{Osamu05,Sota05}.  We also 
showed that the analytical solution in \cite{Evans05} can be characterized as  a 
critical solution connecting two fixed points as is shown in  \cite{Dehnen05}. We got the results that the 
LV relation is rather special in that the critical solution 
connecting the two fixed points exist only in the case with the virial ratio
$b=1$ among the 
models with general constant $b$ models. 
Here we investigate whether or not the model with the LV relation 
can explain several characters of the cosmological simulations by using
the constant $\beta$ critical solutions. 

\section{\label{sec:secII} Analysis and conclusion}

\begin{figure}[h]
\includegraphics[width=24pc]{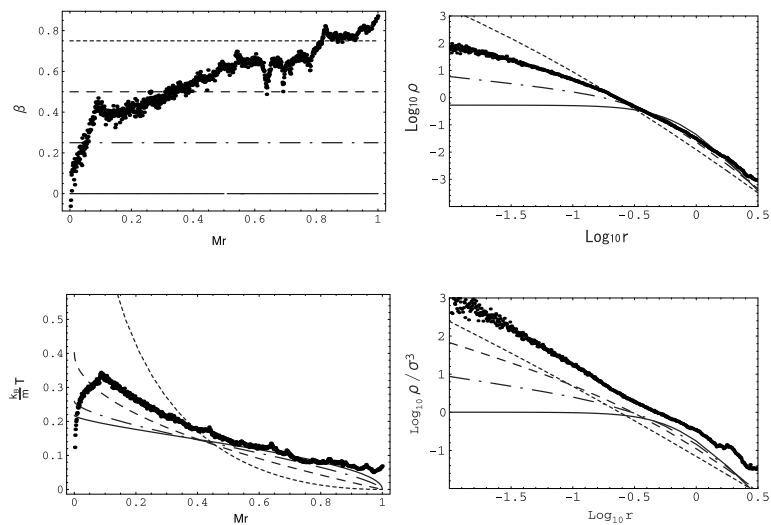}\hspace{2pc}%
\begin{minipage}[b]{12pc}
    \caption{ \label{figcosmo}The behaviour of several physical quantities for a cosmological simulation. In the left figures, the horizontal axis represents the cumulative mass normalized with the
total mass of the bound state.
Each line represents the critical
solution
with $\beta=0$(solid line),
$0.25$(dot-dashed line), $0.5$(dashed line), $0.75$(dotted line), respectively.The density fits well to  $\beta=0.5$ critical solution, while phase space density
fits rather to $\beta=0.75$ critical solution. 
}
\end{minipage}
\end{figure}

Here we examined the character of the bound state for a typical
cosmological simulation. 
The numerical data of the cosmological simulation  was provided to us by H.Kase
at Tokyo University.
We got the result that the LV  relation is satisfied quite well
in the cosmological simulation except for the inner region of a  bound
state. 
We
examined the character of the bound state derived from  the
cosmological simulation
with the constant $\beta$ critical solutions under the LV relation. 
The difference of the results
between cold collapse  and cosmological simulations are
remarkable. In cold
collapse simulations from a homogeneous sphere, 
the central part of the bound state becomes flat, which
can be described by the $\beta $=0 critical solution, while the
outer part can be described by connecting  the  inner Plummer model  with
outer  $\beta >0 $ critical solution \cite{Osamu05,Sota05}. 
In the cosmological simulation, on the other hand, the inner part
is peculiar, i.e., the temperature $k_B T/m(=\sigma^2/3)$ falls  down in the central part of the
bound state (Fig.\ref{figcosmo}). In this region, the LV relation is not attained, while  the phase space density takes a scaling property quite well. 
These results indicate that there are two types of initial
conditions ,i.e., the one settling down to 
the bound state with the LV
relation and the other settling to  the scale invariant  phase-space density.

\end{document}